%% file: trimer.tex
\documentstyle[epsf]{europhys}
\input euromacr

\begin{document}

\euro{}{}{}{}
\Date{}
\shorttitle{J. S\'OLYOM and J. ZITTARTZ: EXACT TRIMER GROUND STATES}

\title{Exact trimer ground states on a spin-1 chain\footnote{Work
performed within the research program of the Sonderforschungsbereich 341, 
K\"oln-Aachen- J\"ulich}}
\author{J. S\'olyom\inst{1}\footnote{E-mail: solyom@mail.szfki.kfki.hu}  
    \And J. Zittartz\inst{2}\footnote{E-mail: zitt@thp.uni-koeln.de}}
\institute{
     \inst{1} Research Institute for Solid State Physics and Optics, 
     H-1525 Budapest, P.O.Box 49, Hungary\\
     \inst{2} Institut f\"ur Theoretische Physik, Universit\"at zu K\"oln,
      Z\"ulpicherstr.\ 77, 50937 K\"oln, Germany}

\rec{}{}

\pacs{
\Pacs{75}{10.Jm}{Quantized spin models}
}

\maketitle

\begin{abstract}
We construct a new spin-1 model on a chain. Its ground state is 
determined exactly which is three-fold degenerate by breaking 
translational invariance. Thus we have trimerization. Excited states cannot
be obtained exactly, but we determine a few low-lying ones by using trial
states, among them solitons.
\end{abstract}

Spin chains and spin ladders have been intensively studied in recent years 
\cite{dagotto}. Besides the Bethe ansatz solvable models there is an 
increasing number of models where at least the ground state is known 
exactly and where eventually the gap to the first excited state can be 
calculated \cite{mikeska}. Typically in the ground state of these models 
the correlations are of short range which implies a finite gap. One natural 
way to construct such models in one dimension is to use the matrix product 
method \cite{zittartz} which is based on the concept of "optimum ground
states" as explained in ref.~\cite{optimum}.

In some cases translational symmetry is spontaneously broken leading to
twofold degenerate, dimerized ground states. To the best knowledge of the 
authors no model has been found so far where the ground state would be
threefold degenerate, thus leading to trimerization. For the first time
we construct such a model.

There exists one model, the Uimin-Lai-Sutherland (ULS) model \cite{uls},
which is a special case of the spin-1 bilinear-biquadratic chain, 
where the excitation spectrum has a tripled periodicity in the Brillouin 
zone, namely soft modes appear at $q=0$ and $q=\pm 2 \pi/3$, however, the 
ground state itself does not break translational symmetry.

For later convenience the Hamiltonian of the spin-1 bilinear-biquadratic model
is written in the form
\begin{equation}
   {\cal H}= \sum_{i=1}^N \left[\left({\tt \bf S}_i \cdot {\tt \bf S}_{i+1} 
    \right)^2
    + \alpha \left({\tt \bf S}_i \cdot {\tt \bf S}_{i+1} \right) \right] .
\end{equation}   
The ULS point corresponds to $\alpha=1$. For $\alpha > 1$ a gap opens up in 
the excitation spectrum above the unique ground state, if the chain is 
periodically closed. At $\alpha=3$ we get the AKLT model \cite{aklt}, where
the existence of the gap has been proven. For $\alpha < 1$ the gapless 
spectrum with three soft modes presumably survives \cite{solyom,itoi}. 
Since a gapless spectrum and, correspondingly, power law like decaying 
correlations should be an exception, and generically a spin chain 
should have a finite gap, one might expect to find a gapped trimerized state
by perturbing away from the Hamiltonian given above. 

To construct such a model that has an exact threefold degenerate ground state
with a finite gap to the excitations we observe that the
quantity
\begin{equation}
   A_{i,j} = \left({\tt \bf S}_i \cdot {\tt \bf S}_{j} \right)^2
    + \alpha \left({\tt \bf S}_i \cdot {\tt \bf S}_{j} \right) +
     \alpha - 1
\end{equation}
has the following properties. As eigenstates arrange in multiplets, we denote 
the 2-spin singlet state at site $i$ and $j$ by $s[i,j]$, and similarly by 
$t[i,j]$ and $q[i,j]$ the triplet and quintuplet states. Then we find
\begin{equation}
     A_{i,j} \left\{ 
       \begin{array}{c}
        s[i,j] \\[1mm]
        t[i,j] \\[1mm]
        q[i,j] \end{array} \right\}  = 
      \left\{ 
       \begin{array}{r}
       (3-\alpha)\,s[i,j]\,, \\[1mm]
       0 \,t[i,j]\,, \\[1mm]
       2 \alpha \,q[i,j] \,.
        \end{array}  \right.
\label{energies}
\end{equation}
This implies that the triplet has the lowest energy in the range 
$0 < \alpha < 3$. 

For three spins at sites $i,j,k$ one has one septuplet, two quintuplets,
three triplets, and precisely one totally antisymmetric singlet. This 
trimer singlet is given by
\begin{equation}
     S[i,j,k]= \frac{1}{\sqrt{6}} \left[ (+,0,-) + (0,-,+) + (-,+,0) - 
      (-,0,+) - (0,+,-) - (+,-,0)  \right] 
\end{equation} 
in terms of the three states $(+)$, $(-)$ and $(0)$ of a spin-1 operator.
Then using the fact that in order to form a singlet with the third spin, any 
two spins have to form a triplet, one gets easily from eq.~(\ref{energies})
\begin{equation}
     A_{i,j} \, S[i,j,k] = A_{i,k} \, S[i,j,k] = A_{j,k} \, S[i,j,k] = 0 .
\end{equation} 

Using this property one sees immediately that the Hamiltonian
\begin{equation}
      {\cal H} = \sum_{i=1}^N h_{i,i+1,i+2,i+3} = \sum_{i=1}^N 
      A_{i,i+2} A_{i+1, i+3}
\label{Hamiltonian}
\end{equation}
has precisely three zero energy ground states. For a finite periodically
closed chain of $N=3 p$ sites, where $p$ is an integer, these three states 
are:
\begin{eqnarray}
  \Psi_1 &=& \prod_{i=1}^p S[3i-2, 3i, 3i+2] \,,  \nonumber \\
  \Psi_2 &=& \prod_{i=1}^p S[3i-1, 3i+1, 3i+3] \,,  \\
  \Psi_3 &=& \prod_{i=1}^p S[3i, 3i+2, 3i+4] \,. \nonumber 
\label{ground}
\end{eqnarray}
They are ground states, as ${\cal H}$ and $h$ of eq.~(\ref{Hamiltonian}) are
positive-semidefinite in $0 < \alpha < 3$, and they are optimum ground states,
as in the sense of ref.~\cite{optimum} they are simultaneously ground 
states of all local interactions $h$. For a finite system these states are 
not strictly orthogonal, but their overlap goes to zero exponentially 
with $N \rightarrow \infty$ namely as $\gamma^N$ ($0 < \gamma < 1$).

\begin{figure}[htb]
\epsfxsize=13truecm \centerline{\epsfbox{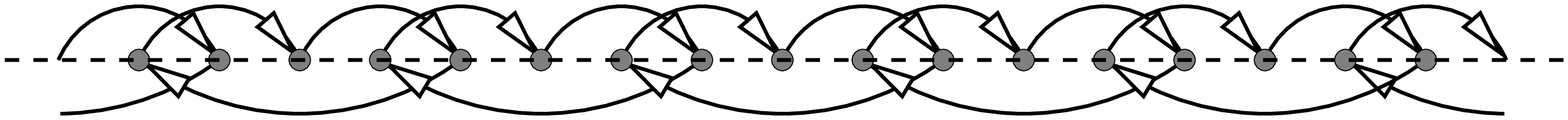}}
\caption{Valence bond representation of the ground state.}
\label{fig1}
\end{figure}

\begin{figure}[htb]
\epsfxsize=13truecm \centerline{\epsfbox{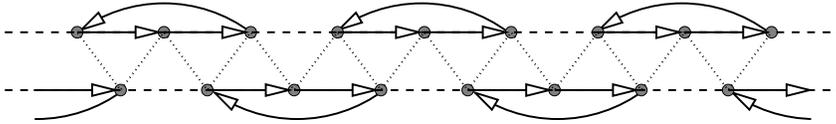}}
\caption{The ground state configuration represented on a zig-zag ladder.}
\label{fig2}
\end{figure}

To visualize the states we show one of them in fig.~{\ref{fig1}} by connecting
the sites belonging to the trimer singlets by valence bonds. Because the 
trimer singlet is antisymmetric under left to right reflection, {\sl i.e.,}
$[i,j,k] \rightarrow [k,j,i]$, the bonds have to be directed. 
It is more convenient to draw the chain as a zig-zag ladder which is shown 
in fig.~\ref{fig2}. When $N=6p$, this is a genuine ladder, while when 
$N=3(2p+1)$, it can be thought of as part of a Moebius ribbon. In what 
follows we will use everywhere this zig-zag ladder representation.   

\vspace{-8mm}

\vspace{-3mm}

That these states are in fact the ground states of the Hamiltonian
in the range $0 < \alpha < 3$, follows immediately from eq.~\ref{energies}. 
At $\alpha=0$ the quintuplet of the spins at site $i$ and $j$ has the same
energy as the singlet, so at this point the degeneracy of the ground state 
becomes exponentially large. For $\alpha < 0$ the quintuplet has the lowest 
energy, leading to the fully polarized ferromagnetic state. The transition
of the trimer phase to the ferromagnetic one is of first order.

At $\alpha = 3$ the 2-spin singlet and triplet local pairs become degenerate,
producing again an exponentially large degeneracy, while for $\alpha > 3$
we get to the massive, Haldane-like regime with unique ground state.

Next we discuss excited states of the model. They cannot be calculated 
exactly, but clearly there has to be a gap to the excitations, because the 
ground state is formed of short range valence bonds. The simplest excited 
state could be a state where one trimer singlet is 
broken up and is replaced by a triplet (Tri), quintuplet (Quint) or 
septuplet (Sept) of the three spins, {\sl e.g.}, 
\begin{equation}
    \Phi_{\rm B}(l) =  B[3l-2, 3l, 3l+2] \, \prod_{i \neq l } 
      S[3i-2, 3i, 3i+2] \,,
\end{equation}
where $B$ stands for Tri, Quint or Sept. Using these as trial states and
calculating their energy as the average $E = \langle {\cal H} \rangle$,
we see that two terms of the Hamiltonian give a non-zero contribution:
\begin{equation}
    A_{3l-2, 3l} A_{3l-1, 3l+1} + A_{3l-1, 3l+1} A_{3l, 3l+2}\,.
\end{equation}
As can be easily seen the Hamiltonian has no matrix element between states 
where different trimers are broken, and there is no overlap between these 
states. Therefore these excited states have no dispersion. For the three 
possible triplet states we get the energy
\begin{equation}
     E_{\rm Tri} = \left\{ \begin{array}{l}
          (2 + \alpha)({1\over 3} + \alpha) \,, \\[1mm]
          (3 -a_1 -\alpha)({1 \over 3} + \alpha) \,,  \\[1mm]   
          (3 -a_2 -\alpha)({1 \over 3} + \alpha) \,,  
\end{array}\right.
\label{tri-ener}
\end{equation} 
where $a_{1,2}$ are the solutions of the equation
\begin{equation}
    a_{1,2}= 1 - {\textstyle{3 \over 2}}\alpha \pm \sqrt{ \left(1 - 
    {\textstyle{3 \over 2}}\alpha\right)^2 + 3 - \alpha } \,.
\end{equation}
For the two quintuplet states the energies are 
\begin{equation}
     E_{\rm Quint} = \left\{ \begin{array}{l}
         \alpha ( {1\over 3} + \alpha) \,, \\[1mm]
          3\alpha ( {1\over 3} + \alpha) \,,  
       \end{array}\right.
\label{quint-ener}
\end{equation} 
while for the septuplet state $E_{\rm Sept} = 4\alpha ( 1/3 + \alpha)$.
One checks that the triplet gap with $a_1$ in eq.~(\ref{tri-ener}) and
the first quintuplet gap in eq.~(\ref{quint-ener}) are the lowest ones.

The lowest energy propagating modes are probably solitonic excitations, 
which are moving domain walls between two different ground states.
In a ring of length $N = 3p$, where the lattice spacing is taken to be
unity, a soliton cannot appear alone. In the simplest case one needs three
solitons to satisfy the periodic boundary condition. However, provided
the chain is long enough, solitons can be studied separately assuming
that the two ends are in different ground states. 

A domain wall that perturbes the system as little as possible can be
obtained on a chain of $3p+2$ sites between state $\Psi_1$ on the left 
hand side and $\Psi_3$ on the right hand side by having the spins at 
sites $3l-2$ and $3l+1$ forming a singlet, triplet or quintuplet, while 
all other spins are in their trimer singlet ground state. Similar domain 
walls appear between state $\Psi_2$ on the left and $\Psi_1$ on
the right or $\Psi_3$ on the left and $\Psi_2$ on the right. 

\begin{figure}[htb]
\epsfxsize=13truecm \centerline{\epsfbox{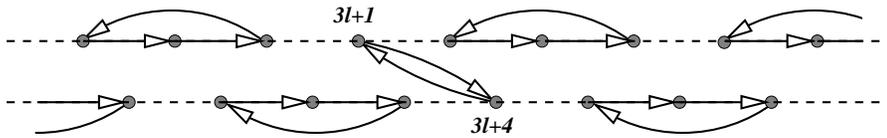}}
\caption{Domain wall on a chain with $3p+2$ sites represented on a 
zig-zag ladder.}
\label{fig3}
\end{figure}

Such a state can be written as
\begin{equation}
  \Phi_{\rm c}(l) = \prod_{i=1}^{l-1} S[3i-2, 3i, 3i+2] \, c[3l-2,3l+1] 
     \, \prod_{j=l}^p S[3j, 3j+2, 3j+4] \,,  \\
\end{equation} 
where $c$ stands for $s$, $t$ or $q$. It is shown in fig.~\ref{fig3}.

\vspace{-3mm}

Only one term in the Hamiltonian will give a non-zero contribution to the
energy, namely
\begin{equation}
    A_{3l-2, 3l} A_{3l-1, 3l+1}\,.
\end{equation}
Although the Hamiltonian has no matrix element between states with different
site index $l$, a propagating mode is obtained since there is a finite 
overlap between these states. Now looking for the moving soliton in the form
\begin{equation}
     \Phi_{\rm c}(q) = \sum_{l=1}^p e^{i 3 q l } \Phi_{\rm c}(l) \,,
\end{equation}
we find for the energy of the singlet and quintuplet excitations
\begin{equation}
     E_{\rm s,q}(q) = {\textstyle{1 \over 4}} \left( {\textstyle{1 \over 3}}
    + \alpha \right)^2 [5 - 3 \cos 3 q], 
\end{equation} 
and 
\begin{equation}
     E_{\rm t}(q) = {\textstyle{1 \over 4}} \left( {\textstyle{1 \over 3}}
    + \alpha \right)^2 [5 + 3 \cos 3 q] 
\end{equation} 
for the triplet excitations.

\begin{figure}[htb]
\epsfxsize=13truecm \centerline{\epsfbox{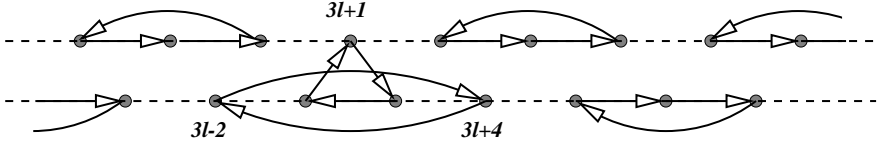}}
\caption{Larger domain wall on a chain with $3p+2$ sites represented on a 
zig-zag ladder.}
\label{fig4}
\end{figure}
\vspace{-3mm}

A somewhat more complicated domain wall is shown in fig.~\ref{fig4}, 
between state $\Psi_1$ on the left and state $\Psi_3$ on the right. 
Because of its symmetric shape, the Hamiltonian has finite matrix elements 
between domain walls on different sites, giving direct propagation. In
addition to that the overlap has also to be considered, giving finally a
rather complicated dispersion relation. When the two spins form a singlet, 
the dispersion relation of the soliton is:
\begin{equation}
   E_{\rm s}(q) = {\textstyle{2 \over 9}} \left[ \left(1 + 3 \alpha \right)^2
    + \left( 1 - 2 \alpha \right)^2 \cos 3 q  \right] { 5 - 3 \cos 3 q  
     \over 14/3 - 2 \cos 3 q }, 
\end{equation}
while for the triplet
\begin{equation}
   E_{\rm t}(q) = 2 \left( {\textstyle{1\over 3}} + \alpha \right)^2 \left[ 
      1 + {\textstyle{1 \over 4}} \cos 3 q  \right] { 5+ 3 \cos 3 q  
      \over 16/3 + 4 \cos 3 q }, 
\end{equation}
and for the quintuplet
\begin{equation}
   E_{\rm q}(q) = {\textstyle{2 \over 9}} \left[ \left(1 + 3 \alpha \right)^2
    + {\textstyle{1 \over 4}} \left( 1 +\alpha \right)^2 \cos 3 q  \right] { 
      5 - 3 \cos 3 q  \over 14/3 - 2 \cos 3 q }. 
\end{equation}

It is interesting to speculate what happens if the model presented in this 
paper is perturbed away from the form given in eq.~(\ref{Hamiltonian}). As 
mentioned earlier both at $\alpha = 0$ and $\alpha=3$ the degeneracy of 
the ground state gets exponentially large. Beyond these points the system 
becomes ferromagnetic, or a first order phase transition to a massive Haldane 
phase occurs. If extra parameters are introduced by making the couplings
of the two-spin and four-spin terms independent, the trimerized phase
will survive because of its finite gap. However, at some point the two legs 
of the zig-zag ladder become decoupled, and other phases may appear. 

Finally we mention that the model can easily be generalized without changing
essential properties. Namely for $N=6p$ one can have different parameters
$\alpha$ and $\beta$ in the operators $A$ acting on the upper and the
lower legs, respectively. We like to thank Dr. A. Schadschneider for this hint.

Furthermore we mention that we have found other model Hamiltonians with 
trimer ground states. However, in all these cases they are degenerate with 
other ground states and the degeneracy is exponentially large. Clearly they
are not so interesting.

\stars

One of the authors (J. S.) is grateful to the University of Cologne for the
hospitality during his visit, where most of this work was done. The authors 
acknowledge the financial support of the Deutsche Forschungsgemeinschaft.
The work was partially supported by the Hungarian Research Fund (OTKA) grant
No. 30173.

\end{document}

%% file: euromacr.tex
\def\And{{\rm and\ }}

\def\stars{\bigskip\centerline{***}\medskip}

\newif\ifboo \boofalse

\def\Review#1{\boofalse{\it #1},}
\def\Name#1{{\sc #1},}
\def\Vol#1{\ifboo Vol. {\bf #1}\else{\bf #1}\fi}
\def\Year#1{\ifboo #1\else(#1)\fi}

\def\Page#1{\ifboo {\rm p. #1}\else{\rm #1}\fi}